\documentclass[Afour,sageh,times,doublespace]{sagej}

\usepackage[colorlinks,bookmarksopen,bookmarksnumbered,citecolor=red,urlcolor=red]{hyperref}
\usepackage{xurl}

\usepackage{graphicx}%
\usepackage{enumitem}%
\usepackage{algorithm}%
\usepackage{algorithmic}%

\DeclareMathOperator{\trunc}{trunc}
\DeclareMathOperator{\diag}{diag}
\DeclareMathOperator{\rmod}{rmod}
\DeclareMathOperator{\round}{round}
\DeclareMathOperator{\single}{single}
\DeclareMathOperator{\rand}{rand}
\DeclareMathOperator{\randn}{randn}

\DeclareMathOperator{\logf}{\_\_log2f}

\newtheorem{theorem}{Theorem}

\newcommand\BibTeX{{\rmfamily B\kern-.05em \textsc{i\kern-.025em b}\kern-.08em
T\kern-.1667em\lower.7ex\hbox{E}\kern-.125emX}}

\def\volumeyear{2016}

\begin{document}

\runninghead{Kawakami and Takahashi}

\title{Improved Scaling for Fast Mode of Ozaki Scheme II}

\author{Shota Kawakami\affilnum{1} and Daisuke Takahashi\affilnum{2}}

\affiliation{\affilnum{1}Graduate School of Science and Technology, University of Tsukuba, Japan\\
\affilnum{2}Center for Computational Sciences, University of Tsukuba, Japan}

\corrauth{Shota Kawakami, Graduate School of Science and Technology, University of Tsukuba
Tennodai~1-1-1, Tsukuba, Ibaraki, 305-8573, Japan}

\email{kawakami@hpcs.cs.tsukuba.ac.jp}

\begin{abstract}
Ozaki scheme II emulates high-precision matrix multiplication using low-precision integer matrix operations based on the Chinese remainder theorem (CRT).
It first scales the high-precision matrices to convert them into integer matrices.
For this scaling step, Ozaki scheme II provides two modes: accurate mode, which uses INT8 matrix multiplication to estimate scaling factors, and fast mode, which applies the Cauchy--Schwarz inequality at lower computational cost.
We show that the existing formula lacks scale invariance; multiplying the input matrices by a constant changes the effective bit width of the integer matrices in the scaling step, causing accuracy degradation or CRT recovery failure.
To address this, we propose a revised scaling formula derived from the CRT uniqueness condition via the Cauchy--Schwarz inequality.
The proposed formula is scale-invariant by construction, guarantees that the CRT uniqueness condition is always satisfied, and introduces no additional overhead over the original fast mode.
Experiments on an NVIDIA GH200 GPU show that the proposed method achieves accuracy comparable to that of accurate mode while maintaining throughput comparable to that of fast mode.
In the accuracy--throughput trade-off, the proposed method overcomes the accuracy limitation of fast mode and the throughput constraint of accurate mode, offering a superior accuracy and performance.
\end{abstract}


\maketitle

\section{Introduction}
Modern accelerators such as graphics processing units (GPUs) have been designed with increasingly powerful low-precision arithmetic units, driven by growing demand from the artificial intelligence community. 
For example, on the NVIDIA GH200 GPU, the FP64 Tensor Core achieves 67~TFLOPS (tera floating-point operations per second) and the INT8/FP8 Tensor Cores achieve 1979~TOPS (tera operations per second)/TFLOPS~\citep{NVIDIA2025}.
On the more recent NVIDIA B200 GPU, the performance gap is wider: the FP64 Tensor Core achieves 37~TFLOPS and the INT8/FP8 Tensor Cores achieve 4500~TOPS/TFLOPS~\citep{NVIDIA-Blackwell-Arch}.

Scientific computing requires high-precision arithmetic such as FP64 or FP32.
Methods that emulate high-precision computation using low-precision hardware, as well as mixed-precision algorithms that exploit low-precision operations for part of the computation, have thus attracted considerable attention~\citep{Dongarra2024}.
Among them, the Ozaki scheme~\citep{Ozaki2012,Ozaki2025} has emerged as a promising approach.

The Ozaki scheme emulates high-precision matrix multiplication using low-precision matrix operations.
When there is a large performance gap between low- and high-precision units, exploiting the high-throughput low-precision hardware enables fast and energy-efficient high-precision matrix multiplication.
\cite{Mukunoki2020} were the first to implement single- and double-precision general matrix multiply (SGEMM and DGEMM, respectively) using the Ozaki scheme with FP16 Tensor Cores, which are low-precision matrix computation units.
\cite{Ootomo2024} implemented DGEMM using even lower-precision INT8 Tensor Cores.
More recently, \cite{Mukunoki2026} proposed an implementation of DGEMM using FP8 Tensor Cores.
These implementations achieve performance that exceeds that of cuBLAS on NVIDIA GPUs.
Since the core of the implementations relies on BLAS (Basic Linear Algebra Subprograms), no highly specialized implementation or hand-tuned optimization is required, making the approach highly portable.
The Ozaki scheme has also been applied to matrix multiplication in scientific computing applications, including quantum circuit simulation~\citep{Ootomo2024}, quantum chemistry calculations~\citep{Dawson2024}, and ab initio tensor network state methods~\citep{Brower2026}.

Recently, Ozaki scheme II~\citep{Ozaki2025} was proposed as an improvement over the conventional Ozaki scheme~\citep{Ozaki2012}.
Ozaki scheme II converts high-precision matrices into integer matrices and decomposes the computation into a series of low-precision integer matrix multiplications via the Chinese remainder theorem (CRT).
Compared to the conventional Ozaki scheme, Ozaki scheme II improves performance by reducing the number of low-precision matrix multiplications required.
On the NVIDIA GH200 GPU, cuBLAS DGEMM achieves a peak performance of 60.9~TFLOPS, whereas DGEMM based on Ozaki scheme II reaches up to 80.2~TFLOPS~\citep{Ozaki2025}.
Uchino et al.\ proposed methods for implementing DGEMM and SGEMM via Ozaki scheme II using INT8 matrix engines~\citep{Uchino2025a} and FP8 matrix engines~\citep{Uchino2026a}, and extended the approach to single- and double-precision complex general matrix multiply (CGEMM and ZGEMM, respectively)~\citep{Uchino2025}.
These methods are available in the open-source library GEMMul8~\citep{Uchino2026b}.

In Ozaki scheme II, the high-precision matrices are first scaled to convert them into integer matrices.
During this scaling process, the scaling factors must be chosen to satisfy the conditions for correct CRT recovery.
Ozaki scheme II provides two modes for computing the scaling factors: accurate mode, which uses INT8 matrix multiplication to estimate scaling factors, and fast mode, which applies the Cauchy--Schwarz inequality at lower computational cost.
The trade-off between accuracy and computation time is different between these two modes.
A performance evaluation of the two modes is provided by \cite{Ozaki2025} and the implementation details are described by \cite{Uchino2025a}.

In this paper, we propose an improved fast mode scaling formula for DGEMM and SGEMM via Ozaki scheme II with INT8 matrix engines that achieves better accuracy without sacrificing throughput.
The main contributions of this paper are as follows.

\begin{itemize}
    \item We show that the existing scaling formula for fast mode lacks scale invariance, which can cause accuracy degradation or even CRT recovery failure.
    \item We propose an improved scaling formula for fast mode that addresses this limitation.
    \item We compare the accuracy, throughput, and their trade-off of the proposed method against accurate mode, the original fast mode, and cuBLAS on an NVIDIA GH200 GPU.
\end{itemize}

Throughout this paper, we focus on DGEMM and SGEMM using INT8 matrix engines.
However, since an analogous scaling formula is used in the fast mode of DGEMM and SGEMM with FP8 matrix engines and of ZGEMM and CGEMM with INT8 matrix engines, the proposed improvement is applicable to these cases as well.

The rest of this paper is organized as follows.
Section~\ref{sec:preliminaries} introduces notations and provides an overview of Ozaki scheme II and its two scaling modes.
Section~\ref{sec:improvement} analyzes the fast mode scaling formula and identifies its limitation through theoretical analysis and numerical demonstration.
Section~\ref{sec:proposed} proposes an improved scaling formula that addresses this limitation.
Section~\ref{sec:evaluation} evaluates the proposed method in terms of accuracy, throughput, and their trade-off, comparing it against accurate mode, the original fast mode, and cuBLAS.
Section~\ref{sec:conclusion} concludes the paper.

\section{Preliminaries}
\label{sec:preliminaries}
\subsection{Notations}
We follow the notations of~\cite{Uchino2025a}.
$\mathbb{F}$ denotes a floating-point number system, such as $\mathbb{F}_{32}$ for FP32 and $\mathbb{F}_{64}$ for FP64.
The operations $\bmod$ and $\rmod\left(\cdot, \cdot\right)$ are respectively defined as
\begin{align*}
    x \bmod p&:= x - p \cdot \left\lfloor x/p \right\rfloor, \\
    \rmod\left(x, p\right) &:= x - p \cdot \round\left(x/p\right),
\end{align*}
where $\round(\cdot)$ denotes rounding to the nearest integer with ties broken by round-to-nearest-even.
$\trunc(\cdot)$ denotes truncation toward zero.
$\diag(\cdot)$ constructs a diagonal matrix from a vector.
$\single(\cdot)$ evaluates its argument in single precision with round-to-nearest-even rounding, while $\single_{\bigtriangleup}(\cdot)$ and $\single_{\bigtriangledown}(\cdot)$ denote single-precision evaluation with round-up and rounding-down, respectively.
When any of these functions is applied to a matrix, it acts element-wise.
$a_{i,:}$ and $b_{:,j}$ denote the $i$-th row of $A$ and the $j$-th column of $B$, respectively.
Throughout this paper, we assume that an INT8 matrix engine accepts two INT8 matrices as inputs and accumulates the inner products in INT32.

\subsection{Ozaki Scheme II}
Ozaki scheme II~\citep{Ozaki2025} is a high-precision matrix multiplication emulation algorithm based on the following CRT, using low-precision matrix multiplication.

\begin{theorem}[Chinese remainder theorem]
    Let $p_1, \ldots, p_{N} \in \mathbb{N}_{\geq 2}$ be pairwise coprime integers,
    and let $P := \prod_{i=1}^{N} p_i$.
    For $x \in \mathbb{Z}$, we define
    \begin{equation*}
        x_i := x \bmod p_i \quad (i=1, \ldots, N).
    \end{equation*}
    Let $P_i := P / p_i$ and let $q_i$ be the modular inverse of $P_i$ modulo $p_i$.
    Then, the following holds:
    \begin{equation*}
        x \equiv \sum_{i=1}^{N} P_i q_i x_i \bmod{P}.
    \end{equation*}
\end{theorem}

Ozaki scheme II computes the product $C \approx AB \in \mathbb{F}^{m \times n}$ of matrices $A = (a_{ij}) \in \mathbb{F}^{m \times k}$ and $B = (b_{ij}) \in \mathbb{F}^{k \times n}$ via the CRT in the following steps.

\begin{enumerate}[label=\textbf{Step \arabic*.}, leftmargin=*, align=left]
    \item Choose pairwise coprime integers $p_1, \ldots, p_{N} \in \mathbb{N}_{\geq 2}$. Compute $P := \prod_{i=1}^{N} p_i$, $P_i := P / p_i$, and $q_i := P_i^{-1} \bmod p_i$.

    \item Using $\mu = \left(\mu_i\right) \in \mathbb{F}^{m}$ and $\nu = \left(\nu_j\right) \in \mathbb{F}^{n}$, convert $A$ and $B$ into integer matrices $A' = (a'_{ij}) \in \mathbb{Z}^{m \times k}$ and $B' = (b'_{ij}) \in \mathbb{Z}^{k \times n}$, respectively, as follows:
    \begin{align*}
        a'_{ij} &:= \trunc\left( \mu_i a_{ij} \right), \\
        b'_{ij} &:= \trunc\left(  \nu_j b_{ij}\right).
    \end{align*}
    Here, $\mu$ and $\nu$ are chosen such that $A'$ and $B'$ satisfy the following condition:
    \begin{align}
         \left|\sum_{h = 1}^{k} a'_{ih} b'_{hj}\right| \leq \sum_{h = 1}^{k} \left|a'_{ih}\right| \left|b'_{hj}\right| < \frac{P}{2} \quad \forall i, j. \label{eq:os2-condition}
    \end{align}

    \item Compute $C'' \equiv A'B' \bmod{P}$ using the CRT as follows:
    \begin{align}
        C' &:= \sum_{i=1}^{N} P_i q_i \cdot \rmod\left(A', p_i\right) \cdot \rmod\left(B', p_i\right), \label{eq:os2-C'}\\
        C'' &:= \rmod\left(C', P\right). \label{eq:os2-C''}
    \end{align}

    \item Rescale $C''$ using $\mu$ and $\nu$ to obtain $C \approx AB$ as follows:
    \begin{align*}
        C &:= \diag\left(\mu\right)^{-1} \cdot C'' \cdot \diag\left(\nu\right)^{-1}.
    \end{align*}
\end{enumerate}

By choosing $p_i$ as positive integers representable in INT8, the matrix products $\rmod\left(A', p_i\right) \cdot \rmod\left(B', p_i\right)$ in \eqref{eq:os2-C'} can be computed using an INT8 matrix engine, provided that $k \leq 2^{17}$ so that the INT32 accumulator does not overflow; for $k > 2^{17}$, block matrix multiplication can be applied to avoid overflow.

By choosing $\mu$ and $\nu$ to satisfy \eqref{eq:os2-condition}, $A'B'$ can be exactly recovered in \eqref{eq:os2-C''}.
The bit width of $A'$ and $B'$ affects the accuracy of $C$; increasing it requires increasing the number of moduli $N$, and hence enlarging $P$.

The dominant cost is Step 3, which performs INT8 general matrix multiply (GEMM) on $A'$ and $B'$ for each modulus in \eqref{eq:os2-C'}.
Since the $O(mnk)$ INT8 GEMM is performed $N$ times (once per modulus), the total computational cost is $O(Nmnk)$.

\subsection{Scaling in Ozaki Scheme II}
\label{sec:gemmul8}

Ozaki scheme II provides two modes for determining $\mu$ and $\nu$: accurate mode and fast mode.
The two modes differ in how they evaluate condition~\eqref{eq:os2-condition}.

In accurate mode, an auxiliary INT8 GEMM is used to evaluate condition~\eqref{eq:os2-condition}.
First, we define $\mu' = \left(\mu'_i\right) \in \mathbb{F}^m$ and $\nu' = \left(\nu'_j\right) \in \mathbb{F}^n$ from $A$ and $B$, respectively, as follows:
\begin{align*}
    \mu'_i &:= 2^{5 - \left\lfloor \log_2 \left(\max_{1 \leqq h \leqq k} |a_{ih}|\right) \right\rfloor}, \\
    \nu'_j &:= 2^{5 - \left\lfloor \log_2 \left(\max_{1 \leqq h \leqq k} |b_{hj}|\right) \right\rfloor}.
\end{align*}
Here, $\left\lfloor \log_2(\cdot) \right\rfloor$ can be computed directly from the exponent field of a floating-point number.
Then, we convert $A$ and $B$ into INT8 matrices $\bar{A} = \left(\bar{a}_{ij}\right) \in \mathbb{Z}^{m \times k}$ and $\bar{B} = \left(\bar{b}_{ij}\right) \in \mathbb{Z}^{k \times n}$ using $\mu'$ and $\nu'$, respectively, as follows:
\begin{align*}
    \bar{a}_{ij} &:= \left\lceil \mu'_i \left|a_{ij}\right| \right\rceil \leq 2^7 - 1, \\
    \bar{b}_{ij} &:= \left\lceil \nu'_j \left|b_{ij}\right| \right\rceil \leq 2^7 - 1.
\end{align*}
Using $\bar{C} := \bar{A}\bar{B} = (\bar{c}_{ij}) \in \mathbb{Z}^{m \times n}$, which can be computed via an INT8 matrix engine, $\mu$ and $\nu$ are defined as follows:
\begin{align}
    \mu_i &:= \mu^{\prime}_i \cdot 2^{\left\lfloor P'_{\mathrm{accu}} - 0.5 \cdot \log_2 \left(\max_{1 \leqq h \leqq n} \bar{c}_{ih}\right) \right\rfloor}, \label{eq:mu_accu}\\
    \nu_j &:= \nu^{\prime}_j \cdot 2^{\left\lfloor P'_{\mathrm{accu}} - 0.5 \cdot \log_2 \left(\max_{1 \leqq h \leqq m} \bar{c}_{hj}\right) \right\rfloor}. \label{eq:nu_accu}
\end{align}
Here, $P'_{\mathrm{accu}} \in \mathbb{F}_{32}$ is a precomputed single-precision constant defined as
\begin{equation*}
    P'_{\mathrm{accu}} := \single_{\bigtriangledown}\left(\log_2(P - 1)/2 - 0.5\right).
    \label{eq:P-prime-accu}
\end{equation*}
The $\log_2(\cdot)$ in \eqref{eq:mu_accu} and~\eqref{eq:nu_accu} is evaluated using the single-precision fast base-2 logarithm function $\logf$ provided by the CUDA Math API on NVIDIA GPUs~\citep{CUDAMathAPI} or the HIP Math API on AMD GPUs~\citep{HIPMathAPI}.
The $\logf$ in the CUDA Math API has an absolute error of at most $4u_{32}$ for inputs $x \in [0.5, 2]$ and an absolute error of at most 2~ULP (unit in the last place) for other inputs~\citep{CUDAMathAPI}.
For the HIP Math API, the absolute error is at most 1~ULP for $x \in [10^{-6}, 10^{6}]$~\citep{HIPMathAPI}.
According to \cite{Uchino2025}, normalizing $x > 0$ to $x' := x / 2^{\left\lfloor \log_2(x) \right\rfloor} \in [1, 2)$ yields
$\logf(x') \leq \log_2(x') / (1 - 4u_{32})$.
Thus, defining
\begin{equation}
    \delta := \single_{\bigtriangleup}\left(\frac{0.5}{1 - 4u_{32}}\right),
    \label{eq:delta}
\end{equation}
and evaluating $\log_2(x)$ as
\begin{equation}
    \widetilde{\log_2}(x) := \single_{\bigtriangleup}\left(\logf\left(\frac{x}{2^{\left\lfloor \log_2(x) \right\rfloor}}\right) + \left\lfloor \log_2(x) \right\rfloor\right),
    \label{eq:flog2}
\end{equation}
\eqref{eq:mu_accu} and~\eqref{eq:nu_accu} can be computed as follows:
\begin{align}
    \mu_i &:= \mu^{\prime -1}_i \cdot 2^{\left\lfloor P'_{\mathrm{accu}} - \delta \cdot \widetilde{\log_2} \left(\max_{1 \leqq h \leqq n} \bar{c}_{ih}\right) \right\rfloor}, \\
    \nu_j &:= \nu^{\prime -1}_j \cdot 2^{\left\lfloor P'_{\mathrm{accu}} - \delta \cdot \widetilde{\log_2} \left(\max_{1 \leqq h \leqq m} \bar{c}_{hj}\right) \right\rfloor}.
\end{align}
With $\mu$ and $\nu$ defined as above, $A'$ and $B'$ satisfy condition~\eqref{eq:os2-condition} as follows:
\begin{align*}
    2 \sum_{h = 1}^{k} \left|a'_{ih}\right| \left|b'_{hj}\right|
    &\leq 2 \mu_i \left(\sum_{h = 1}^{k}|a_{ih}| |b_{hj}|\right) \nu_j \\
    &\leq 2 \mu_i \mu^{\prime -1}_i \bar{c}_{ij} \nu_j \nu^{\prime -1}_j < P.
\end{align*}
The dominant cost in computing $\mu$ and $\nu$ is the auxiliary INT8 GEMM, which has a time complexity of $O(mnk)$.
A detailed error analysis of accurate mode is provided by~\cite{Uchino2026}.

In fast mode, the estimation is performed using the Cauchy--Schwarz inequality.
That is, $\mu$ and $\nu$ are chosen to satisfy the following:
\begin{align}
    2 \sum_{h = 1}^{k} \left|a'_{ih}\right| \left|b'_{hj}\right| &\leq 2 \mu_i \left\|a_{i,:}\right\|_2 \left\|b_{:,j}\right\|_2 \nu_j < P. \label{eq:fast-mode-condition}
\end{align}
$\mu$ and $\nu$ are defined by the following formulas, where $\widetilde{\log_2}(\cdot)$ and $\delta$ are as given in \eqref{eq:flog2} and~\eqref{eq:delta}, respectively:
\begin{align}
    \mu_i &:= 2^{\left\lfloor P'_{\mathrm{fast}} - \max\left(1, \delta\cdot\widetilde{\log_2}\left(\sum_{h=1}^{k} a_{ih}^2 \right)\right) \right\rfloor - \left\lfloor \log_2 \left(\max_{1 \leq h \leq k} |a_{ih}| \right) \right\rfloor}, \label{eq:fast-mode-mu}\\
    \nu_j &:= 2^{\left\lfloor P'_{\mathrm{fast}} - \max\left(1, \delta\cdot\widetilde{\log_2} \left(\sum_{h=1}^{k} b_{hj}^2 \right)\right) \right\rfloor - \left\lfloor \log_2 \left(\max_{1 \leq h \leq k} |b_{hj}| \right) \right\rfloor}. \label{eq:fast-mode-nu}
\end{align}
Here, $P'_{\mathrm{fast}} \in \mathbb{F}_{32}$ is a precomputed single-precision constant defined as
\begin{equation*}
    P'_{\mathrm{fast}} := \single_{\bigtriangledown}\left(\log_2(P - 1)/2 - 1.5\right).
    \label{eq:P-prime-fast}
\end{equation*}
\begin{algorithm}[tb]
\caption{Computation of $\mu_i$ in fast mode (row $i$ of $A$)}
\label{alg:fast-mode-mu}
\begin{algorithmic}[1]
\REQUIRE $a_{\mathrm{norm}} := \left(\sum_{h=1}^{k} a_{ih}^2\right)$ computed with rounding-up
\REQUIRE $a_{\mathrm{max}} := \max_{1 \leq h \leq k} |a_{ih}|$
\REQUIRE $P'_{\mathrm{fast}} := \single_{\bigtriangledown}\left(\log_2(P - 1)/2 - 1.5\right) \in \mathbb{F}_{32}$
\REQUIRE $\delta := \single_{\bigtriangleup}\left(0.5 / (1 - 4u_{32})\right) \in \mathbb{F}_{32}$
\ENSURE $\log_2 \left(\mu_i\right) \in \mathbb{Z}$
\STATE $s := \left\lfloor \log_2\left(a_\mathrm{norm}\right) \right\rfloor$
\STATE $t := \single_{\bigtriangleup}\left(\delta \cdot \left(\logf\left(a_{\mathrm{norm}} / 2^{s}\right) + s\right)\right)$
\STATE return $\left\lfloor P'_{\mathrm{fast}} - \max\left(1.0, t\right) \right\rfloor -  \left\lfloor \log_2 \left(a_\mathrm{max}\right) \right\rfloor$
\end{algorithmic}
\end{algorithm}
In \eqref{eq:fast-mode-mu} and~\eqref{eq:fast-mode-nu}, the sums $\sum_{h=1}^{k} a_{ih}^2$ and $\sum_{h=1}^{k} b_{hj}^2$, respectively, are evaluated in single or double precision with rounding-up; these sums, together with the row- and column-wise maximum values, dominate the cost of computing $\mu$ and $\nu$, requiring $O(mk)$ and $O(kn)$ operations, respectively.
The detailed computation of $\mu_i$ is summarized in Algorithm~\ref{alg:fast-mode-mu}.

Accurate mode provides a tighter estimate, allowing a larger bit width for $A'$ and $B'$ than that in fast mode.
However, it incurs an additional computational cost for the auxiliary INT8 GEMM.
Fast mode avoids this overhead by using the Cauchy--Schwarz inequality, which is less expensive to evaluate, resulting in shorter computation time than that of accurate mode.
However, since the estimate is looser, the bit width of $A'$ and $B'$ may be smaller than that in accurate mode, potentially reducing accuracy.

\section{Limitation of Fast Mode Scaling}
\label{sec:improvement}
Here, we identify a potential limitation in the fast mode scaling formulas~\eqref{eq:fast-mode-mu}.
We analytically show that \eqref{eq:fast-mode-mu} lacks scale invariance and may cause CRT recovery failure.
We then present the results of numerical experiments to demonstrate that these issues occur in practice.

\subsection{Theoretical Analysis}
\label{sec:limitation}
In the scaling step (Step~2) of Ozaki scheme~II, the $i$-th row of the scaled integer matrix $A'$ has an effective precision of
\begin{equation*}
    \log_2 \left(\mu_i\right) + \left\lfloor \log_2 \left(\max_{1 \leq h \leq k} |a_{ih}| \right) \right\rfloor + 1
\end{equation*}
bits. When fast mode is used, this precision is evaluated as follows:
\begin{align*}
    &\log_2 \left(\mu_i\right) + \left\lfloor \log_2 \left(\max_{1 \leq h \leq k} |a_{ih}| \right) \right\rfloor + 1 \nonumber \\
    =& \left\lfloor P'_{\mathrm{fast}} - \max\left(1, \delta\cdot\widetilde{\log_2}\left(\sum_{h=1}^{k} a_{ih}^2 \right)\right) \right\rfloor \\
    &- \left\lfloor \log_2 \left(\max_{1 \leq h \leq k} |a_{ih}| \right) \right\rfloor + \left\lfloor \log_2 \left(\max_{1 \leq h \leq k} |a_{ih}| \right) \right\rfloor + 1 \nonumber \\
    =& \left\lfloor P'_{\mathrm{fast}} - \max\left(1, \delta\cdot\widetilde{\log_2}\left(\sum_{h=1}^{k} a_{ih}^2 \right)\right) \right\rfloor + 1.
\end{align*}

Let $A = \alpha \hat{A}$ for some scalar $\alpha = 2^s$ ($s \in \mathbb{Z}$) and matrix $\hat{A} \in \mathbb{F}^{m \times k}$.
Ideally, the scaled integer matrix $A' = \trunc(\diag(\mu) A)$ should be invariant with respect to $\alpha$; that is, its entries and bit width should be independent of $\alpha$.
This is because, to maximize accuracy, $\mu$ should be chosen to maximize the bit width of the scaled integer matrix $A'$ subject to the CRT condition~\eqref{eq:os2-condition}.
More precisely, for $\alpha_0 = 2^{s_0}$ and $\alpha_1 = 2^{s_1}$ ($s_0, s_1 \in \mathbb{Z}$), let $A_0 = \alpha_0 \hat{A}$ and $A_1 = \alpha_1 \hat{A}$, and let $\mu_0$ and $\mu_1$ be the scaling factors determined by \eqref{eq:fast-mode-mu} for $A_0$ and $A_1$, respectively.
Then, ideally $A'_0 = \trunc\!\left(\diag(\mu_0) A_0\right)$ and $A'_1 = \trunc\!\left(\diag(\mu_1) A_1\right)$ should be equal.
In fast mode, however, this scale invariance does not hold.

For $A = \alpha\hat{A}$, the effective precision of the $i$-th row of $A'$ then becomes:
\begin{align}
    &\left\lfloor P'_{\mathrm{fast}} - \max\left(1, \delta\cdot \left(\widetilde{\log_2}\left(\sum_{h=1}^{k} \hat{a}_{ih}^2 \right) + 2 \log_2 (\alpha) \right) \right) \right\rfloor \nonumber \\
    &+ 1. \label{eq:effective-precision-fast}
\end{align}
When 
\begin{equation*}
1 < \delta\cdot \left(\widetilde{\log_2}\left(\sum_{h=1}^{k} \hat{a}_{ih}^2 \right) + 2 \log_2 (\alpha) \right),  
\end{equation*}
this expression depends on $\alpha$, and the effective precision of $A'$ decreases as $\alpha$ increases; hence, the fast mode scaling lacks scale invariance.
\begin{figure}[tb]
    \centering
    \includegraphics[width=1.0\hsize]{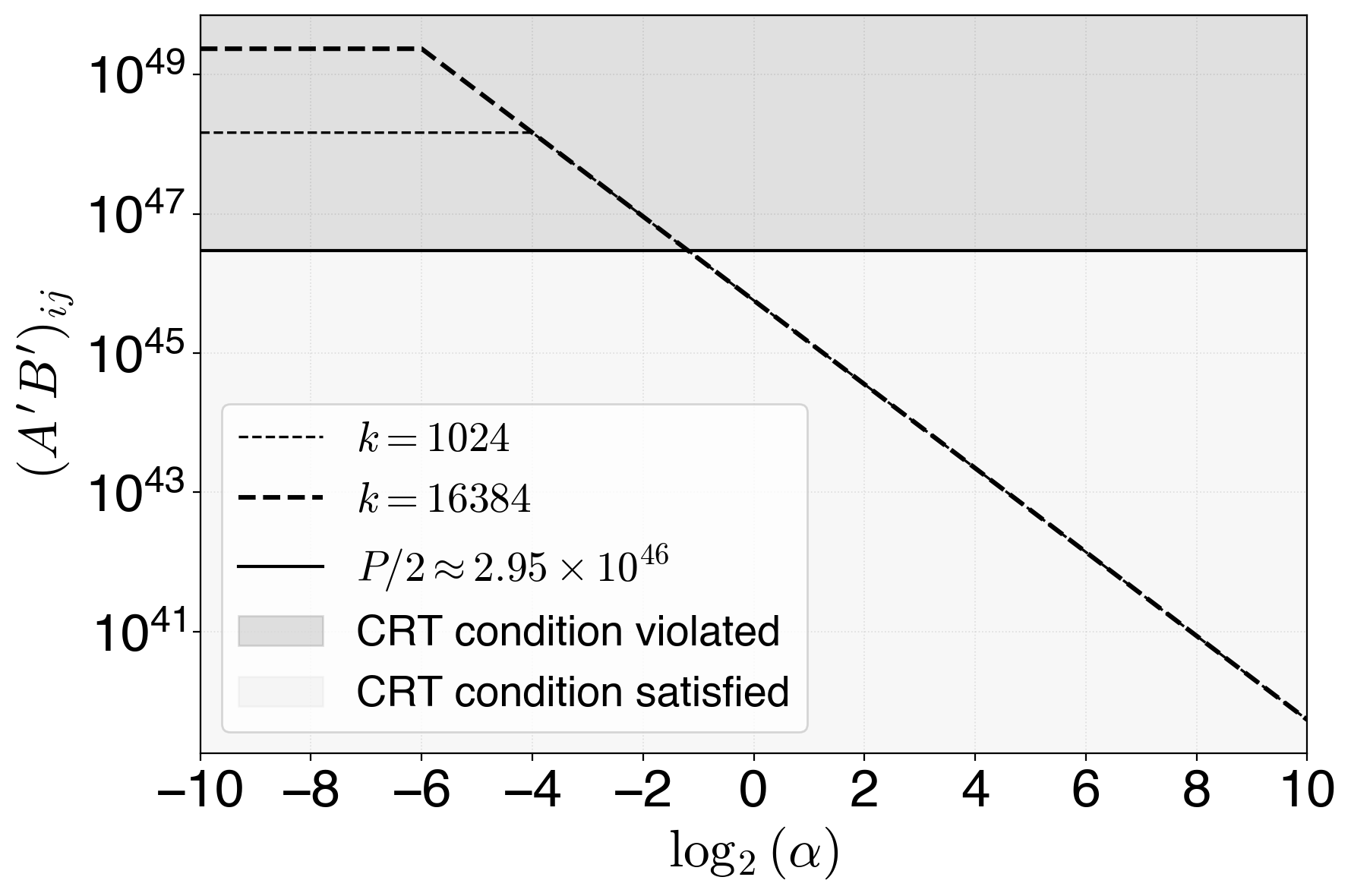}
    \caption{$(A'B')_{ij}$ in fast mode versus scalar $\alpha = 2^s$, where $A = \alpha\hat{A}$ and $B = \alpha\hat{B}$ with $\hat{a}_{ij} = \hat{b}_{ij} = 1$ (all-ones matrices), and $N = 20$ moduli.
    The horizontal black line marks the CRT recovery threshold $P/2$.
    The dark-shaded region indicates where the CRT uniqueness condition is violated ($(A'B')_{ij} > P/2$), and the light-shaded region indicates where it is satisfied.
    Thin and thick lines correspond to $k = 1024$ and $k = 16384$, respectively, where $k$ is the inner dimension of the matrix product.
    }
    \label{fig:mu_condition}
\end{figure}

On the other hand, when 
\begin{equation*}
1 \geq \delta\cdot \left(\widetilde{\log_2}\left(\sum_{h=1}^{k} \hat{a}_{ih}^2 \right) + 2 \log_2 (\alpha) \right),
\end{equation*}
the effective precision does not depend on $\alpha$ and is fixed at $\left\lfloor P'_{\mathrm{fast}} - 1 \right\rfloor + 1$ bits; however, $A'B'$ may exceed $P/2$, causing the CRT recovery to fail.
For example, suppose $\hat{a}_{ij} = 1$ for all $i, j$ and $\alpha \leq k^{-1/2}$.
Then:
\begin{align*}
    &\delta\cdot \left(\widetilde{\log_2}\left(\sum_{h=1}^{k} \hat{a}_{ih}^2 \right) + 2 \log_2 (\alpha) \right) \\
    &= \delta\cdot \left(\log_2(k) + 2 \log_2 (\alpha) \right) \\
    &\leq \delta \cdot \left(\log_2 (k) - \log_2 (k)\right) = 0 < 1,
\end{align*}
so the scaled integer matrix $A' = \trunc\!\left(\diag(\mu) A\right)$ does not depend on $\alpha$, and its bit width is fixed at $\left\lfloor P'_{\mathrm{fast}} - 1 \right\rfloor + 1$ bits.
Similarly, let $\hat{B} = \left(\hat{b}_{ij}\right) \in \mathbb{F}^{k \times n}$ with $\hat{b}_{ij} = 1$ and $B = \alpha \hat{B}$; then, for $\alpha \leq k^{-1/2}$, $B' = \trunc\!\left(B \diag(\nu)\right)$ is also $\alpha$-independent with the same fixed bit width.
In this case, since $\max_{1 \leq h \leq k} |a_{ih}| = \max_{1 \leq h \leq k} |b_{hj}| = \alpha$ for all $i, j$, the scaling factors $\mu$ and $\nu$ obtained from $A$ and $B$ are given by
\begin{equation*}
    \mu_i = \nu_j = 2^{\left\lfloor P'_{\mathrm{fast}} - 1 \right\rfloor - \log_2(\alpha)} = 2^{\left\lfloor P'_{\mathrm{fast}} - 1 \right\rfloor} \cdot \alpha^{-1}.
\end{equation*}
As all entries of $A$ and $B$ are $\alpha$, all entries of $A' = \trunc\!\left(\diag(\mu) A\right)$ and $B' = \trunc\!\left(B \diag(\nu)\right)$ become $2^{\left\lfloor P'_{\mathrm{fast}} - 1 \right\rfloor}$.
Consequently, each entry of $A'B'$ is $k \cdot 2^{2\left\lfloor P'_{\mathrm{fast}} - 1 \right\rfloor}$.
Since $P$ and $P'_{\mathrm{fast}}$ do not depend on $k$, for sufficiently large $k$, the entries of $A'B'$ exceed $P/2$ and CRT recovery fails.
\begin{figure*}[tb]
    \centering
    \includegraphics[width=1.0\hsize]{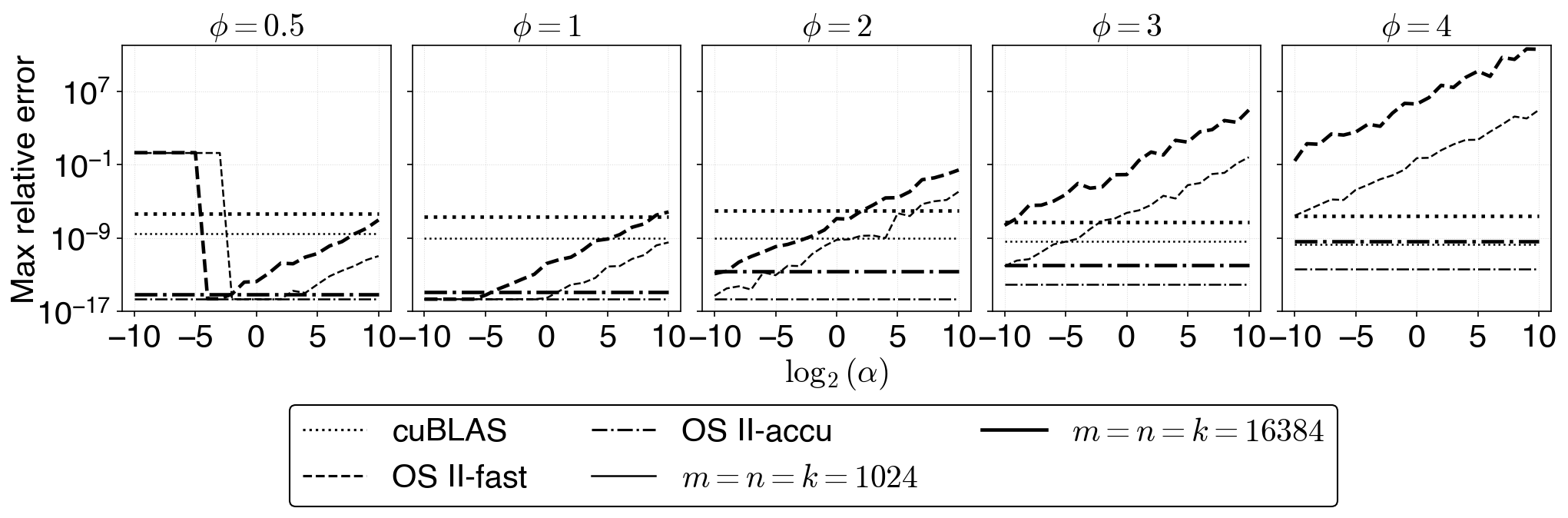}
    \caption{Maximum relative error of DGEMM with respect to double-double precision versus scalar $\alpha = 2^s$ for cuBLAS, OS~II-fast, and OS~II-accu, where $A$ and $B$ are the input matrices given by $A = \alpha\hat{A}$ and $B = \alpha\hat{B}$, with $\hat{A}$ and $\hat{B}$ random matrices generated via \eqref{eq:random-matrix-generation}. Each column corresponds to a different value of $\phi$ controlling the spread of element magnitudes. Matrix dimensions are $m = n = k \in \{1024, 16384\}$ and $N = 20$ moduli. Thin and thick lines correspond to $m = n = k = 1024$ and $m = n = k = 16384$, respectively.}
    \label{fig:accuracy-random}
\end{figure*}

Here, for $k \in \{1024, 16384\}$, we set $\hat{a}_{ij} = 1$ and $\hat{b}_{ij} = 1$, and evaluate $(A'B')_{ij}$ for input matrices $A = \alpha\hat{A}$ and $B = \alpha\hat{B}$ with $\alpha = 2^s$ ($s = -10, -9, \ldots, 10$); the results are shown in Fig.~\ref{fig:mu_condition}.
We use $N = 20$ moduli as defined in GEMMul8:
\begin{align}
    \{p_i\}_{i=1}^{N} = \{&256, 255, 253, 251, 247, \nonumber \\
    &241, 239, 233, 229, 227, \nonumber \\
                          &223, 217, 211, 199, 197, \nonumber \\
                          &193, 191, 181, 179, 173\}.
    \label{eq:moduli}
\end{align}
As shown in Fig.~\ref{fig:mu_condition}, $(A'B')_{ij}$ varies with $\alpha$ and exceeds the CRT recovery threshold $P/2$ in the region $\alpha \leq 2^{-2}$.

\subsection{Numerical Demonstration}
\label{sec:numerical-demo}

We verify the above behavior using numerical examples.
All experiments were conducted on an NVIDIA GH200 GPU with CUDA Toolkit 13.2.

We generate double-precision random matrices $\hat{A} = \left(\hat{a}_{ij}\right) \in \mathbb{F}^{m \times k}_{64}$ and $\hat{B} = \left(\hat{b}_{ij}\right) \in \mathbb{F}^{k \times n}_{64}$ according to
\begin{align}
    \hat{a}_{ij},\ \hat{b}_{ij} = \left(\rand - 0.5\right) \cdot \exp\left(\phi \cdot \randn\right), \label{eq:random-matrix-generation}
\end{align}
where $\rand$ denotes a uniform random variable on $[0, 1)$, $\randn$ denotes a standard normal random variable, and $\phi$ is a parameter that controls the spread of the entry magnitudes.
We evaluate $\phi \in \{0.5, 1, 2, 3, 4\}$.
Fixing the matrices $\hat{A}$ and $\hat{B}$, we generate the inputs $A = \alpha \hat{A}$ and $B = \alpha \hat{B}$ for $\alpha = 2^s$, $s \in \{-10, -9, \ldots, 10\}$.
The matrix dimensions are set to $m = n = k \in \{1024, 16384\}$.
We evaluate the maximum relative error with respect to the result computed in double-double precision.

We compare the following methods:
\begin{itemize}
    \item cuBLAS: DGEMM computed by NVIDIA cuBLAS.
    \item OS~II-fast: DGEMM computed by Ozaki scheme II with fast mode implemented in GEMMul8.
    \item OS~II-accu: DGEMM computed by Ozaki scheme II with accurate mode implemented in GEMMul8.
\end{itemize}
The number of moduli used in OS~II-fast and OS~II-accu is set to $N = 20$, with the moduli $\{p_i\}$ as defined in \eqref{eq:moduli}.
Note that for the scaling procedure in GEMMul8 v3.0.4 (the latest version at the time of writing), fast mode differs from Algorithm~\ref{alg:fast-mode-mu}.
However, throughout all evaluations in this paper, to discuss the results based on~\cite{Uchino2025a}, we reverted the part to the procedure in Algorithm~\ref{alg:fast-mode-mu}, which corresponds to the implementation in GEMMul8 v2.0.19.
We confirmed that the fast mode in GEMMul8 v3.0.4 exhibits worse accuracy than that of the fast mode based on Algorithm~\ref{alg:fast-mode-mu} and that it shows accuracy characteristics similar to those of OS~II-fast, as described below.
The modified source code used in this evaluation is publicly available \citep{Kawakami2026}.
\begin{figure*}[tb]
    \centering
    \includegraphics[width=0.6\hsize]{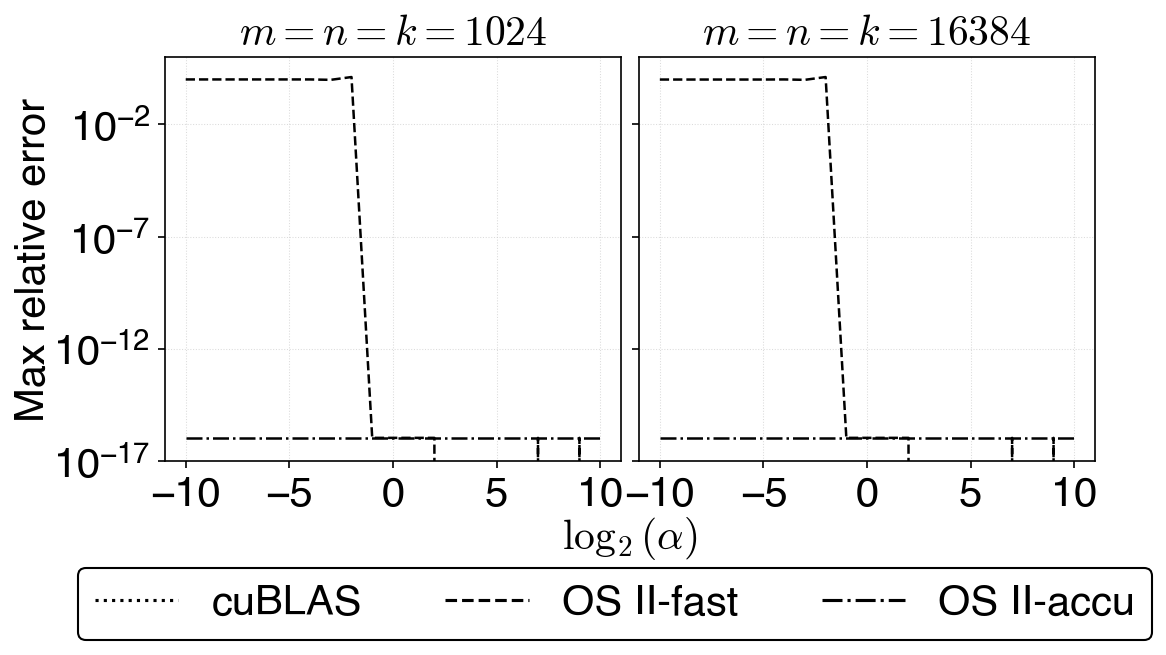}
    \caption{Maximum relative error of DGEMM with respect to double-double precision versus scalar $\alpha = 2^s$ for cuBLAS, OS~II-fast, and OS~II-accu, where $A$ and $B$ are the input matrices given by $A = \alpha\hat{A}$ and $B = \alpha\hat{B}$, with $\hat{a}_{ij} = \hat{b}_{ij} = 1$ (all-ones matrices), $m = n = k \in \{1024, 16384\}$, and $N = 20$ moduli. cuBLAS achieves a maximum relative error of exactly~0 in all cases.}
    \label{fig:accuracy-ones}
\end{figure*}

The results are shown in Fig.~\ref{fig:accuracy-random}.
Both cuBLAS and OS~II-accu maintained constant accuracy across all values of $\alpha$ for each $\phi$ value.
In contrast, for OS~II-fast, accuracy decreased as $\alpha$ increased.
This is consistent with the analysis above; in the regime where
\begin{equation*}
  1 < \delta \cdot \left(\widetilde{\log_2}\!\left(\sum_{h=1}^{k} \hat{a}_{ih}^2 \right) + 2\log_2(\alpha)\right)
\end{equation*}
or
\begin{equation*}
  \quad 1 < \delta \cdot \left(\widetilde{\log_2}\!\left(\sum_{h=1}^{k} \hat{b}_{hj}^2 \right) + 2\log_2(\alpha)\right),
\end{equation*}
the effective bit width of $A'$ or $B'$ depends on $\alpha$ and decreases as $\alpha$ increases.
Furthermore, for $\phi = 0.5$, the maximum relative error reaches approximately~1 in the region $\alpha \leq 2^{-3}$ for $k = 1024$ and $\alpha \leq 2^{-5}$ for $k = 16384$, indicating a complete breakdown of the computation.
This is likely because the entries of the scaled integer matrix product $A'B'$ exceed $P/2$, causing CRT recovery to fail.

In addition, we evaluated the all-ones case ($\hat{a}_{ij} = \hat{b}_{ij} = 1$).
The results are shown in Fig.~\ref{fig:accuracy-ones}.
For cuBLAS, the maximum relative error was exactly~0 in all cases.
For both $k = 1024$ and $k = 16384$, OS~II-accu maintained constant error, whereas OS~II-fast reached a maximum relative error of approximately~1 in the region $\alpha \leq 2^{-2}$, indicating a complete breakdown of the computation.
This is consistent with the results shown in Fig.~\ref{fig:mu_condition}; in the region $\alpha \leq 2^{-2}$, the entries of the scaled integer matrix product $A'B'$ exceed $P/2$, causing CRT recovery to fail.

These observations indicate that the fast mode scaling lacks scale invariance.
In this paper, we propose revised definitions of $\mu$ and $\nu$ that still rely on the Cauchy--Schwarz inequality and introduce no additional computational cost, while ensuring scale invariance.

\section{Proposed Method}
\label{sec:proposed}
We define $\mu$ and $\nu$ directly from the Cauchy--Schwarz bound \eqref{eq:fast-mode-condition}.
With the requirement of
\begin{equation*}
    2 \mu_i \left\|a_{i, :}\right\|_2 \left\|b_{:, j}\right\|_2 \nu_j \leq P - 1 < P \label{eq:proposed-condition-raw}
\end{equation*}
to hold for all $i, j$, a sufficient condition is
\begin{align}
    \mu_i \leq \sqrt{\frac{P - 1}{2}} \cdot \frac{1}{\left\|a_{i, :}\right\|_2}, \quad \nu_j \leq \sqrt{\frac{P - 1}{2}} \cdot \frac{1}{\left\|b_{:, j}\right\|_2}. \label{eq:proposed-condition}
\end{align}
Taking the base-2 logarithm of the bound on $\mu_i$ in \eqref{eq:proposed-condition} gives
\begin{align*}
    \log_2 \left(\mu_i\right) &\leq \log_2 \left(\sqrt{\frac{P - 1}{2}} \cdot \frac{1}{\left\|a_{i, :}\right\|_2}\right) \\ 
    &= \log_2(P-1)/2 - 0.5 - 0.5 \cdot \log_2 \left(\sum_{h = 1}^{k} a_{ih}^2\right).
\end{align*}
We therefore define $\mu_i$ as the largest power of two satisfying this bound:
\begin{align}
    \mu_i &:= 2^{\left\lfloor P'_{\mathrm{prop}} - 0.5 \cdot \left(\widetilde{\log_2}\left(\sum_{h = 1}^{k} a_{ih}^2\right) + 4u_{32}\right)\right\rfloor} \nonumber  \\
    &\leq \sqrt{\frac{P - 1}{2}} \cdot \frac{1}{\left\|a_{i, :}\right\|_2}, \label{eq:proposed-mu}
\end{align}
and analogously for $\nu_j$:
\begin{align}
    \nu_j &:= 2^{\left\lfloor P'_{\mathrm{prop}} - 0.5 \cdot \left(\widetilde{\log_2}\left(\sum_{h = 1}^{k} b_{hj}^2\right) + 4u_{32}\right)\right\rfloor} \nonumber \\
    &\leq \sqrt{\frac{P - 1}{2}} \cdot \frac{1}{\left\|b_{:, j}\right\|_2}. \label{eq:proposed-nu}
\end{align}
Here, $P'_{\mathrm{prop}}$ is defined as
\begin{equation*}
    P'_{\mathrm{prop}} := \single_{\bigtriangledown}\left(\log_2(P - 1)/2 - 0.5\right). \label{eq:P-prime-proposed}
\end{equation*}
The sums $\sum_{h=1}^{k} a_{ih}^2$ and $\sum_{h=1}^{k} b_{hj}^2$ are evaluated in single or double precision with rounding-up.
The existing fast and accurate modes of GEMMul8 bound the error of $\logf$ for $x \in [1, 2)$ as $\logf(x) \leq \log_2(x) / (1 - 4u_{32})$, treating the error as multiplicative.
However, since $\logf$ has an absolute error of at most $4u_{32}$ for $x \in [1, 2)$, the correct bound is $|\logf(x) - \log_2(x)| \leq 4u_{32}$, which gives
\begin{equation*}
    \log_2(x) \leq \logf(x) + 4u_{32}. \label{eq:logf-bound}
\end{equation*}
Accordingly, \eqref{eq:proposed-mu} and~\eqref{eq:proposed-nu} use 
\begin{align*}
    \log_2(x) &\leq \logf\left(x / 2^{\left\lfloor\log_2(x)\right\rfloor}\right) + \left\lfloor\log_2(x)\right\rfloor + 4u_{32} \nonumber \\ 
    &= \widetilde{\log_2}(x) + 4u_{32}.
\end{align*}
The detailed computation of $\mu_i$ is summarized in Algorithm~\ref{alg:prop-mu}.
\begin{algorithm}[tb]
\caption{Computation of $\mu_i$ in proposed method (row $i$ of $A$)}
\label{alg:prop-mu}
\begin{algorithmic}[1]
\REQUIRE $a_{\mathrm{norm}} := \left(\sum_{h=1}^{k} a_{ih}^2\right)$ computed with rounding-up
\REQUIRE $P'_{\mathrm{prop}} := \single_{\bigtriangledown}\left(\log_2(P - 1)/2 - 0.5\right) \in \mathbb{F}_{32}$
\ENSURE $\log_2 \left(\mu_i\right) \in \mathbb{Z}$
\STATE $s := \left\lfloor\log_2(a_{\mathrm{norm}})\right\rfloor$
\STATE $t := \single_{\bigtriangleup}\left(0.5\cdot\logf\left(a_{\mathrm{norm}} / 2^{s}\right) + 2u_{32}\right)$
\STATE $w := \single_{\bigtriangledown}\left(P'_{\mathrm{prop}} - 0.5 s\right)$
\STATE return $\left\lfloor w - t \right\rfloor$
\end{algorithmic}
\end{algorithm}

Using this $\mu_i$, the effective bit width of the $i$-th row of the scaled integer matrix $A' = \trunc\!\left(\diag(\mu) A\right)$ is:
\begin{align*}
    &\log_2 \left(\mu_i\right) + \left\lfloor\log_2\left(\max_{1 \leq h \leq k} \left|a_{ih}\right|\right)\right\rfloor + 1 \nonumber \\
    =& \left\lfloor P'_{\mathrm{prop}} - 0.5 \cdot \left(\widetilde{\log_2}\left(\sum_{h = 1}^{k} a_{ih}^2\right) + 4u_{32} \right) \right\rfloor \nonumber\\
    &+ \left\lfloor\log_2\left(\max_{1 \leq h \leq k} \left|a_{ih}\right|\right)\right\rfloor + 1.
\end{align*}
As in the analysis in Section~\ref{sec:limitation}, by setting $A = \alpha\hat{A}$ ($\alpha = 2^s$, $s \in \mathbb{Z}$), the effective bit width becomes:
\begin{align*}
    &\left\lfloor P'_{\mathrm{prop}} - 0.5 \cdot \left(\widetilde{\log_2}\left(\sum_{h = 1}^{k} \hat{a}_{ih}^2\right) + 2 \log_2 (\alpha) + 4u_{32} \right)\right\rfloor \nonumber \\
    &+\left\lfloor\log_2\left(\max_{1 \leq h \leq k} \left|\hat{a}_{ih}\right|\right) + \log_2 (\alpha)\right\rfloor  + 1 \\
    =& \left\lfloor P'_{\mathrm{prop}} - 0.5 \cdot \left(\widetilde{\log_2}\left(\sum_{h = 1}^{k} \hat{a}_{ih}^2\right) + 4u_{32} \right)\right\rfloor - \log_2 (\alpha)\nonumber \\
    &+\left\lfloor\log_2\left(\max_{1 \leq h \leq k} \left|\hat{a}_{ih}\right|\right)\right\rfloor + \log_2 (\alpha) + 1 \\
    =& \left\lfloor P'_{\mathrm{prop}} - 0.5 \cdot \left(\widetilde{\log_2}\left(\sum_{h = 1}^{k} \hat{a}_{ih}^2\right) + 4u_{32} \right)\right\rfloor \\
    &+ \left\lfloor\log_2\left(\max_{1 \leq h \leq k} \left|\hat{a}_{ih}\right|\right)\right\rfloor + 1. 
\end{align*}
Here, the second equality uses $\lfloor x \pm s \rfloor = \lfloor x \rfloor \pm s$, which holds since $\log_2(\alpha) = \log_2\left(2^s\right) = s \in \mathbb{Z}$.
As $\log_2\left(\alpha\right)$ cancels out in this way, the effective bit width of the scaled integer matrix is constant and independent of $\alpha$.
Furthermore, since $\mu_i$ is designed to satisfy \eqref{eq:proposed-condition}, the entries of the scaled integer matrix product $A'B'$ are always bounded by $P/2$, ensuring that CRT recovery never fails.

The time complexity of computing $\mu$ and $\nu$ is $O(mk)$ and $O(kn)$, respectively, which is the same order as that of fast mode, introducing no additional computational cost and remaining faster than accurate mode.

\section{Evaluation}
\label{sec:evaluation}
We compare the following implementations for both DGEMM and SGEMM in terms of numerical accuracy and computational performance.
\begin{itemize}
    \item cuBLAS: computed by NVIDIA cuBLAS.
    \item OS~II-fast: computed by Ozaki scheme II with fast mode implemented in GEMMul8.
    \item OS~II-accu: computed by Ozaki scheme II with accurate mode implemented in GEMMul8.
    \item OS~II-prop: computed by Ozaki scheme II with fast mode implemented in GEMMul8, with the scaling formula replaced by the proposed one.
\end{itemize}
The source code of OS~II-prop is available in~\citep{Kawakami2026}.
As noted in Section~\ref{sec:numerical-demo}, the scaling procedure of OS~II-fast was reverted to the v2.0.19 implementation (the procedure in Algorithm~\ref{alg:fast-mode-mu}) throughout all evaluations in this section.
All experiments were conducted on an NVIDIA GH200 GPU with CUDA Toolkit 13.2, as in Section~\ref{sec:numerical-demo}.

\subsection{Accuracy}
Input matrices $A \in \mathbb{F}^{m \times k}$ and $B \in \mathbb{F}^{k \times n}$ were generated using \eqref{eq:random-matrix-generation} and the matrix product was computed by each method.
The maximum relative error with respect to a double-double precision reference was measured.
The matrix dimensions were set to $m = n = k \in \{1024, 16384\}$.
The parameter $\phi$ in \eqref{eq:random-matrix-generation} was set to $\phi \in \{0.5, 1, 2, 3, 4\}$ for DGEMM and $\phi \in \{0, 0.5, 1, 1.5\}$ for SGEMM.
For OS~II-fast, OS~II-accu, and OS~II-prop, the number of moduli $N$ was varied over $9 \leq N \leq 20$ for DGEMM and $3 \leq N \leq 12$ for SGEMM.

\begin{figure*}[tb]
    \centering
    \includegraphics[width=1.0\hsize]{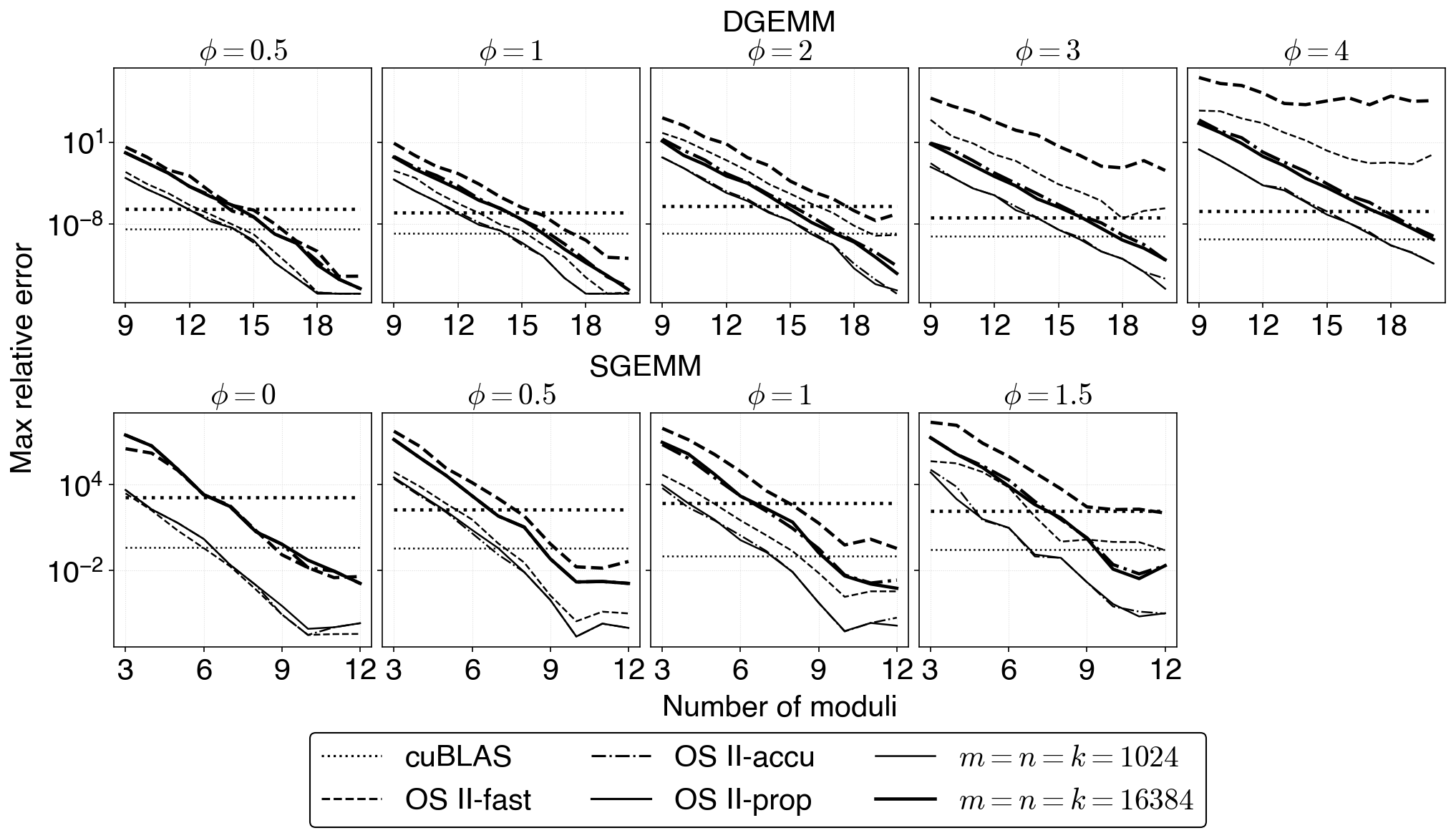}
    \caption{Maximum relative error with respect to double-double precision of DGEMM (top) and SGEMM (bottom) versus number of moduli $N$ for cuBLAS, OS~II-fast, OS~II-accu, and OS~II-prop, using random matrices with $m = n = k$ and varying $\phi$. Thin and thick lines correspond to $m = n = k = 1024$ and $m = n = k = 16384$, respectively.}
    \label{fig:accuracy-comparison}
\end{figure*}

The results are shown in Fig.~\ref{fig:accuracy-comparison}.
For all values of $\phi$, while the maximum relative errors of OS~II-accu, OS~II-fast, and OS~II-prop exceed that of cuBLAS for small $N$, they tend to decrease as $N$ increases.
A comparison of the maximum relative errors of OS~II-fast and OS~II-accu indicates that the gap is small for small values of $\phi$ but grows larger as $\phi$ increases, for both DGEMM and SGEMM.
In particular, for DGEMM with the ill-conditioned cases $\phi \in \{3, 4\}$, the maximum relative error of OS~II-fast exceeds that of cuBLAS.
In contrast, OS~II-prop maintains accuracy comparable to that of OS~II-accu regardless of the value of $\phi$.
Furthermore, OS~II-prop achieves higher accuracy than that of cuBLAS with a sufficient number of moduli, both for DGEMM with $\phi \in \{3, 4\}$.
Although OS~II-prop determines $\mu$ and $\nu$ using the Cauchy--Schwarz inequality, it maintains accuracy comparable to that of OS~II-accu.
Although OS~II-fast also uses the Cauchy--Schwarz inequality, it exhibits accuracy degradation for large $\phi$, suggesting that the root cause of the degradation is not the use of the Cauchy--Schwarz inequality itself.

\subsection{Throughput}
Input matrices $A \in \mathbb{F}^{m \times k}$ and $B \in \mathbb{F}^{k \times n}$ were generated using \eqref{eq:random-matrix-generation}. Matrix multiplication was performed with each method and the execution time was measured.
The matrix sizes were $m = n = k$, where $k \in \{1024, 2048, 4096, 8192, 16384\}$.
The parameter $\phi$ in \eqref{eq:random-matrix-generation} was set to $\phi \in \{0.5, 4\}$ for DGEMM and $\phi \in \{0.5, 1.5\}$ for SGEMM.
For OS~II-fast, OS~II-accu, and OS~II-prop, the number of moduli was varied over $9 \leq N \leq 20$ for DGEMM and $3 \leq N \leq 12$ for SGEMM.
For each method, warm-up iterations were performed until at least 3 iterations had run and either 3 seconds of total time elapsed or 100 iterations were reached; timed iterations then continued until at least 5 iterations had run and either 12 seconds elapsed or 100 iterations were reached. The median execution time $T_\mathrm{med}$ was used to compute throughput in FLOPS as follows:
\begin{equation}
    \mathrm{FLOPS} := \frac{2 m n k}{T_\mathrm{med}}.
    \label{eq:tflops}
\end{equation}

\begin{figure*}[tb]
    \centering
    \includegraphics[width=1.0\hsize]{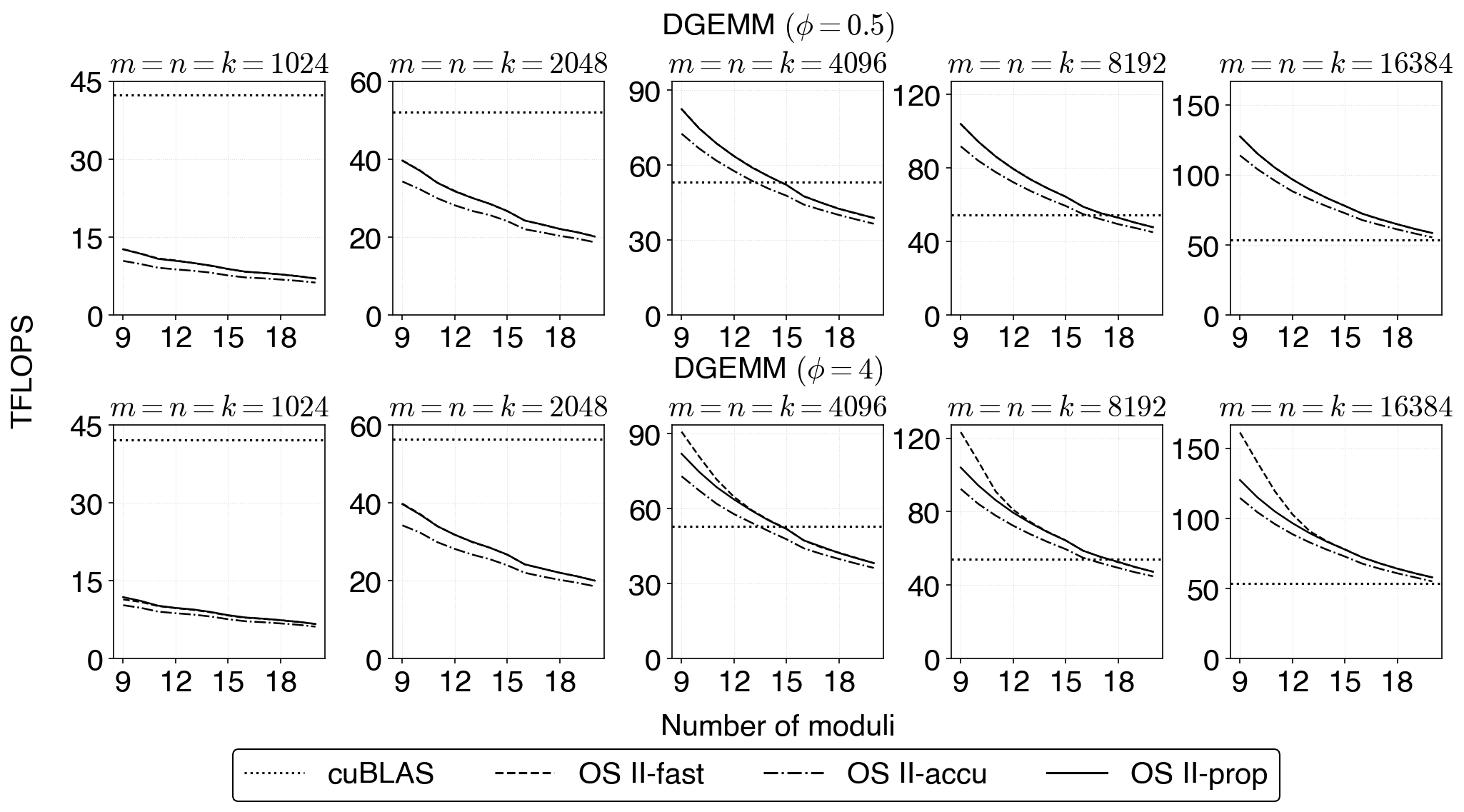}
    \caption{Throughput (TFLOPS) of DGEMM for cuBLAS, OS~II-fast, OS~II-accu, and OS~II-prop versus number of moduli $N$ for random matrices with $\phi = 0.5$ (top) and $\phi = 4$ (bottom), for varying $m = n = k$.}
    \label{fig:tflops-dgemm}
\end{figure*}

\begin{figure*}[tb]
    \centering
    \includegraphics[width=1.0\hsize]{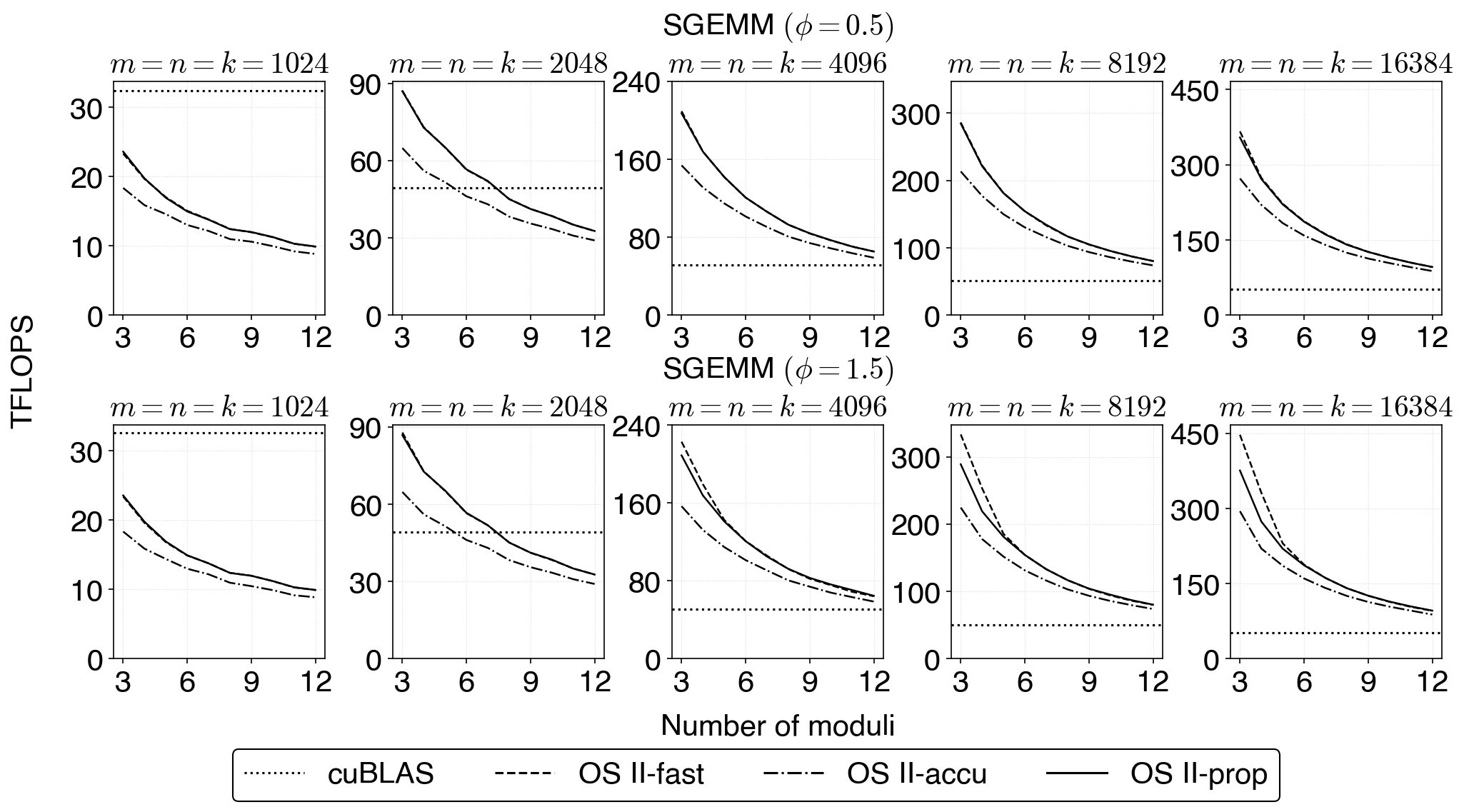}
    \caption{Throughput (TFLOPS) of SGEMM for cuBLAS, OS~II-fast, OS~II-accu, and OS~II-prop versus number of moduli $N$ for random matrices with $\phi = 0.5$ (top) and $\phi = 1.5$ (bottom), for varying $m = n = k$.}
    \label{fig:tflops-sgemm}
\end{figure*}

The results are shown in Figs.~\ref{fig:tflops-dgemm} and~\ref{fig:tflops-sgemm}.
OS~II-prop achieves higher throughput than that of OS~II-accu for both DGEMM and SGEMM.
Since the accuracy evaluation showed that OS~II-prop maintains accuracy comparable to that of OS~II-accu, OS~II-prop delivers higher throughput at equivalent accuracy.

\begin{figure*}[tb]
    \centering
    \includegraphics[width=1.0\hsize]{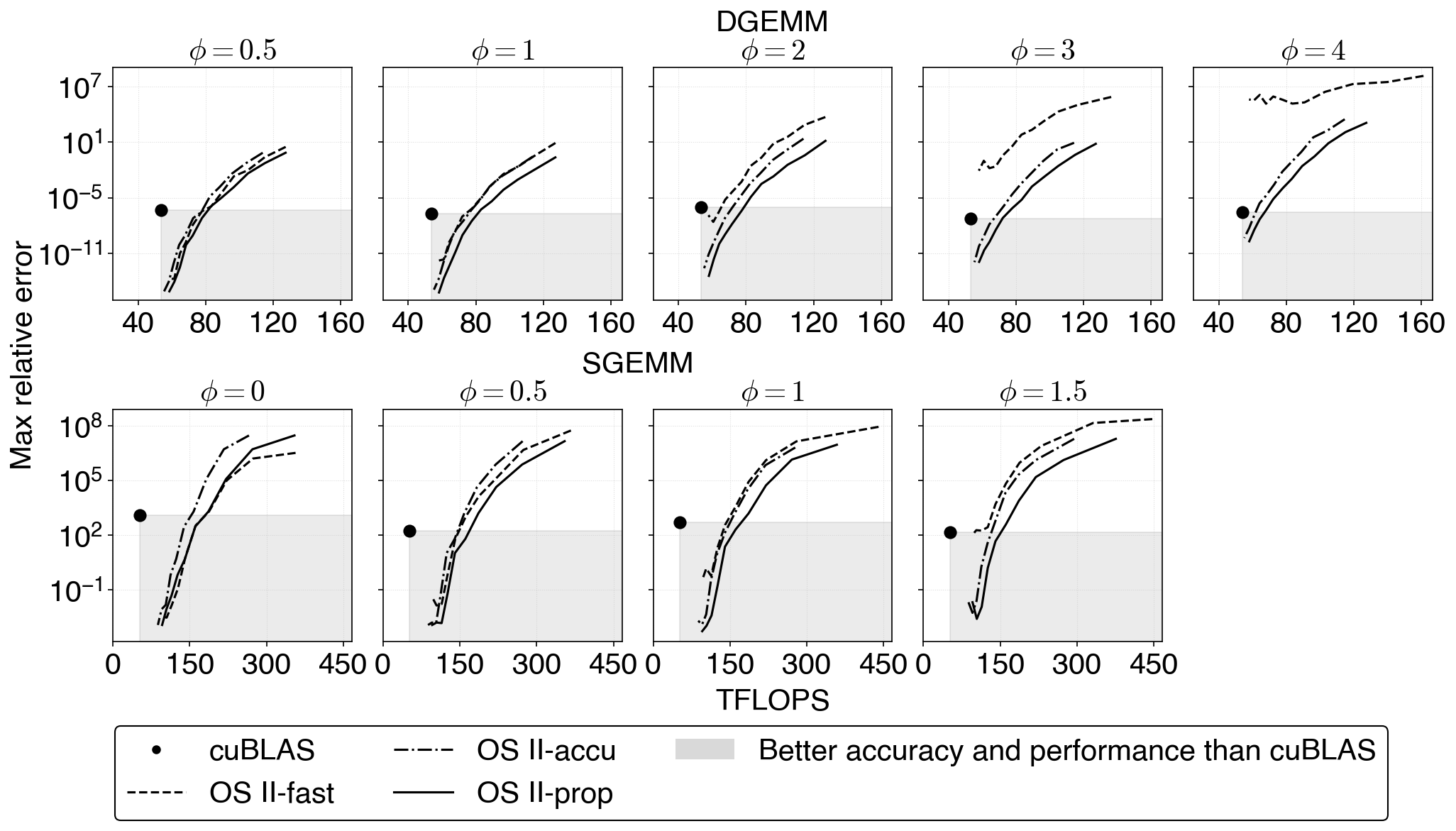}
    \caption{Accuracy--throughput trade-off for DGEMM (top, $\phi \in \{0.5, 1, 2, 3, 4\}$) and SGEMM (bottom, $\phi \in \{0, 0.5, 1, 1.5\}$) at $m = n = k = 16384$.
    The horizontal axis shows throughput in TFLOPS and the vertical axis shows maximum relative error with respect to double-double precision.
    Each curve is traced from right to left as the number of moduli $N$ increases.
    The black dot marks the throughput and accuracy of cuBLAS; the shaded region indicates where a method simultaneously achieves higher accuracy and higher throughput than those of cuBLAS.}
    \label{fig:tradeoff-grid}
\end{figure*}

Both OS~II-prop and OS~II-fast achieve similar throughput for $\phi = 0.5$ across DGEMM and SGEMM.
For $\phi = 4$ in DGEMM and $\phi = 1.5$ in SGEMM, the two methods are also comparable for small matrix sizes and for large matrix sizes with a large number of moduli.
However, for large matrix sizes with a small number of moduli in these cases, OS~II-fast achieves higher throughput than that of OS~II-prop.
We attribute this difference to the fraction of zero elements in the scaled integer matrices.
For $\phi = 0.5$, the spread of element exponents is small. The zero fraction after scaling is therefore similar between OS~II-prop and OS~II-fast.
With a large number of moduli, the CRT bound $P/2$ grows with $N$, allowing a larger range of values to be represented after scaling; consequently, the zero fractions of the two methods become comparable.
With a small number of moduli, however, the representable range is narrower for OS~II-fast due to its smaller scaling factor, resulting in a higher fraction of zero elements.
This effect is especially pronounced for $\phi = 4$ in DGEMM and $\phi = 1.5$ in SGEMM, where the zero fraction of OS~II-fast's scaled integer matrices is substantially higher than that of OS~II-prop.

As noted in the accuracy evaluation, both OS~II-prop and OS~II-fast exhibit considerably higher error than that of cuBLAS when the number of moduli is small.
Therefore, the throughput disadvantage of OS~II-prop at small $N$ is not a practical concern when cuBLAS-level accuracy is required.
The accuracy--throughput trade-off is analyzed in detail in the next subsection.

\subsection{Trade-off between Accuracy and Performance}
To analyze the accuracy--throughput trade-off, we measured additional throughput data for $m = n = k = 16384$, specifically for DGEMM with $\phi \in \{1, 2, 3\}$ and SGEMM with $\phi \in \{0, 1\}$, supplementing the throughput results for DGEMM with $\phi \in \{0.5, 4\}$ and SGEMM with $\phi \in \{0.5, 1.5\}$ shown in Figs.~\ref{fig:tflops-dgemm} and~\ref{fig:tflops-sgemm}; we combined these results with the accuracy data from Fig.~\ref{fig:accuracy-comparison}.

The combined results are shown in Fig.~\ref{fig:tradeoff-grid}, where the horizontal axis represents throughput and the vertical axis represents the maximum relative error.
The black dot indicates the throughput and accuracy of cuBLAS, and the shaded region marks the area where a method achieves both higher accuracy and higher throughput than those of cuBLAS.
For each of OS~II-fast, OS~II-accu, and OS~II-prop, the data points trace a curve from the upper right to the lower left as the number of moduli increases, reflecting the trade-off between accuracy and throughput.

The curve of OS~II-prop lies consistently to the right of that of OS~II-accu for all values of $\phi$.
This indicates that OS~II-prop achieves higher throughput than that of OS~II-accu at any given accuracy level. OS~II-prop therefore dominates OS~II-accu in the accuracy--throughput trade-off.

The curve of OS~II-fast shifts upward (toward higher error) as $\phi$ increases.
In particular, for DGEMM with $\phi \in \{3, 4\}$, the curve of OS~II-fast does not reach the shaded region, meaning that OS~II-fast cannot simultaneously outperform cuBLAS in both accuracy and throughput under these conditions.
In contrast, OS~II-accu and OS~II-prop reach the shaded region for all values of $\phi$, meaning that they can simultaneously outperform cuBLAS in both accuracy and throughput.

With the exception of SGEMM with $\phi = 0$, the curve of OS~II-prop lies furthest to the right among all methods, confirming that OS~II-prop achieves the highest throughput for any target accuracy level.
In the SGEMM case with $\phi = 0$, the spread of element exponents is small and OS~II-fast also achieves comparable accuracy; consequently, the curves of OS~II-fast and OS~II-prop nearly coincide.

In summary, OS~II-prop achieves consistently higher throughput than that of OS~II-accu at equivalent accuracy, while resolving the accuracy degradation of OS~II-fast for large $\phi$ without sacrificing throughput.
The proposed method thus overcomes both the accuracy limitation of OS~II-fast and the throughput constraint of OS~II-accu, offering a superior balance of accuracy and performance.

\section{Conclusion}
\label{sec:conclusion}
In this paper, we proposed an improved fast mode scaling for DGEMM and SGEMM via Ozaki scheme II with INT8 matrix engines, achieving better accuracy without sacrificing throughput.

We identified a potential limitation in the scaling formula of the existing fast mode.
We showed that the formula lacks scale invariance: when the input matrix is scaled by a factor $\alpha$, the effective bit width of the scaled integer matrix varies with $\alpha$, leading to accuracy degradation for large $\alpha$ and CRT recovery failure for small $\alpha$.

To address this limitation, we proposed a revised scaling formula derived directly from the CRT uniqueness condition via the Cauchy--Schwarz inequality.
The proposed formula is scale-invariant by construction and guarantees that the entries of the scaled integer matrix product $A'B'$ never exceed the CRT recovery threshold $P/2$.
Its computational complexity is the same as that of the original fast mode, introducing no additional overhead.

Experiments on the NVIDIA GH200 GPU confirmed that the proposed method achieves accuracy comparable to that of accurate mode for all values of $\phi$.
Its throughput also approaches that of the original fast mode as the number of moduli increases.
In the accuracy--throughput trade-off, the proposed method thus overcomes both the accuracy limitation of OS~II-fast and the throughput constraint of OS~II-accu, offering a superior balance of accuracy and performance.

As future work, it would be interesting to extend the proposed scaling formula beyond DGEMM and SGEMM with INT8 matrix engines to complex matrix multiplication (ZGEMM and CGEMM)~\citep{Uchino2025}, as well as to methods that exploit FP8 matrix engines~\citep{Uchino2026b}.
A rigorous error analysis of the proposed method, analogous to that performed for accurate mode in~\citep{Uchino2026}, is also left as future work.

\begin{dci}
  The authors declared no potential conflicts of interest with respect to the research, authorship, and publication of this article.
\end{dci}

\begin{funding}
  This research was supported by JST SPRING (grant number JPMJSP2124) and used the computational resources of Miyabi provided by the Multidisciplinary Cooperative Research Program in the Center for Computational Sciences, University of Tsukuba.
\end{funding}

\begin{sm}
  Not applicable.
\end{sm}

\bibliographystyle{SageH}
\bibliography{IJHPCA2026}

\end{document}